 \documentclass[aps,prd,letterpaper,showpacs,twocolumn,preprintnumbers,floatfix,superscriptaddress]{revtex4}
\usepackage{slashed}
\usepackage{graphicx,color}
\usepackage{epsfig}
\usepackage{subfigure}
\usepackage{epsfig}
\usepackage{dcolumn}
\usepackage{bm}
\usepackage{color}
\usepackage{eufrak} 
\usepackage{amsfonts}

\def\lsim{\mathrel{\rlap{\lower4pt\hbox{\hskip1pt$\sim$}}
    \raise1pt\hbox{$<$}}}         %less than or approx. symbol
\def\gsim{\mathrel{\rlap{\lower4pt\hbox{\hskip1pt$\sim$}}
    \raise1pt\hbox{$>$}}}         %greater than or approx. symbol

\begin{document}

\title{CP Violation at the Finite Temperature}

\author{Wei Chao}
\email{chaowei@bnu.edu.cn}
\affiliation{Center for Advanced Quantum Studies, Department of Physics, Beijing Normal University, Beijing, 100875, China}

\vspace{3cm}

\begin{abstract}

In this letter we explore the spontaneous CP violation at the finite temperature. We show that the CP-violating phase $\varphi$ may only emerge  around the time of the electroweak phase transition, and it can provide a scource for generating the matter-antimatter asymmetry of the universe via the electroweak baryogenesis mechanism (EWBG) if the domain wall relating to $\pm \varphi$ vacuums  is collapsed by a small explicit $Z_2$-breaking term. No extra CP violation is needed! The spontaneous CP  is restored after  the electroweak phase transition, such that there is no constraint  from the electric dipole moment (EDM) measurements. This scenario resolves the tension between the non-observation of EDMs in precision measurement experiments and the requirement of a large CP violation by the EWBG.

\end{abstract}


\maketitle

 \noindent {\bf Introduction} 
The dynamic of the baryon asymmetry of the Universe (BAU) is one of the longstanding problems in particle physics and cosmology.  
Assuming our Universe was matter-antimatter symmetric at its origin, it is reasonable to speculate that interactions involving elementary particles gave birth to the BAU during the subsequent cosmological evolution. 
Combing data of the Plank with that of the WMAP , the observed BAU is~\cite{Ade:2015xua}
\begin{eqnarray}
Y_{\rm obs}\equiv {\rho_B \over s } = (8.61\pm 0.09) \times 10^{-11} \; , \label{observe}
\end{eqnarray}
where $\rho_B$ is the baryon number density, $s$ is the entropy density of the Universe. 
It is well-known that three Sakharov criteria~\cite{Sakharov:1967dj} must be satisfied for a  successful baryogenesis theory: (1) baryon number violation; (2) C and CP violations (CPV); (3) a departure from the thermal equilibrium.  
The Standard Model (SM) itself contains all the necessary ingredients to the realization of baryogenesis. 
However the CPV phase from the CKM mixing matrix in the SM can not give rise to an adequate baryon asymmetry, because QCD damping effects reduce the generated asymmetry  to a negligible amount~\cite{Huet:1994jb}.  
New  physics beyond the SM is needed for a real baryogenesis.

Of various baryogenesis theories, the electroweak baryogenesis (EWBG)~\cite{Kuzmin:1985mm,Rubakov:1996vz,Morrissey:2012db}, is  promising and attractive, because it is testable with a combination of searching for new degrees of freedom at the LHC and low energy examinations of CP invariance at electric dipole moment (EDM) experiments. 
A successful EWBG requires a first order electroweak phase transition (EWPT) and  a large enough CP violation. 
Both requirements can be met in some well-known extensions of the SM, such as minimal  supersymmetric standard model (MSSM)~\cite{Gabbiani:1996hi} and two-Higgs doublet model~\cite{Inoue:2014nva}.  
It is known that CP-violating phases from either  scalar couplings~\cite{Lee:2004we,Chao:2016aer} or  Yukawa couplings~\cite{Chao:2014dpa,Chao:2015uoa,Guo:2016ixx} can give rise to a large enough relaxation coefficient needed for the EWBG. 
However the non-observation of EDMs of atoms, molecules and the electron puts a very strong constraint on the new CP-violating interactions. 
For example, wino induced bargoenesis scenario~\cite{Li:2008kz} was already excluded by the latest ACME result~\cite{Baron:2013eja}.

In this letter we investigate a scenario where the CP was spontaneously broken at the finite temperature and  was recovered after the EWPT.
We study the strength EWPT in this scenario and  explore the possibility of generating an adequate BAU with the same CP phase via the EWBG. 
This scenario reliefs the tension between the non-observation of EDMs and the EWBG requirement for a large CP violation, because there is no spontaneous CP violation in the zero temperature.  
To illustrate our point, the SM is extended with a complex scalar singlet $S$ and a global $U(1)$ symmetry, which is explicitly broken by the Higgs potential.
We focus on the scenario where the universe undergoes a two-step phase transition, i.e.  $\langle H \rangle \neq 0$, $\langle S\rangle =0$ at $T<T_n$, while $\langle H \rangle = 0$, $ |\langle S\rangle| \neq 0$ at $T\in(T_n, ~T_s) $ and the CP is spontaneously broken as $S$ gets non-zero vacuum expectation value (VEV), where $T_n$ is the bubble nucleation temperature and $T_s$ is the temperature of the first-step phase transition. 
As was shown in Ref.~\cite{Espinosa:2011ax,Espinosa:2011eu} there is a large barrier between the electroweak symmetric phase $\langle H \rangle =0$, $\langle S \rangle =v_s e^\varphi$ and the electroweak symmetry broken phase $\langle H \rangle =v_T$, $\langle S \rangle =0$ at the tree-level, so strongly first order EWPT can be easily realized.
To accommodate the EWBG we introduce vector-like top quark $\mathfrak{t}_{L,R}^{}$ with following possible interactions, 
\begin{eqnarray}
\zeta \overline{\mathfrak{t}_L} S t_R^{} + y^\prime_t\overline{Q_L} \tilde H \mathfrak{t}_R + (M_{\mathfrak{t}} + H^\dagger H /\Lambda) \overline{\mathfrak{t}_L} \mathfrak{t}_R^{}  + {\rm h.c.} \label{modell}
\end{eqnarray} 
where $\Lambda$ is the cut off scale and all couplings are real.  
The first term induces CPV interaction  on the bubble wall, which is a seed for a net BAU via the EWBG. 
We find that the observed BAU can be explained  in this scenario if the $Z_2$ symmetry  relating to $S\leftrightarrow -S$  is explicitly broken in the potential, which causes to the collapse of the domain wall. 
No extra CP violation is needed.

The paper is organized as follows: We study the spontaneous CP violation and EWPT in section II and III respectively. Section IV is devoted to the investigation of EWBG in this model. The last part is concluding remarks. 

 \noindent {\bf Spontaneous CP violation at finite T} 
We extend the SM with a complex scalar singlet $S$, vector-like quark $\mathfrak{t}_{L,R}^{}$ and global $U(1)$ symmetry, under which only $S$ and $\mathfrak{t}$ carries non-zero charge. 
It was pointed out in Ref.~\cite{Haber:2012np}  that {\it spontaneous CP violation in the theory of one complex scalar field may occur only when the related $U(1)$ is explicitly broken by at least two spurions whose $U(1)$ charges are different in magnitude}, where spurion refers to the coupling   in the potential that breaks the $U(1)$ explicitly.   
The Higgs potential  can be written as 
\begin{eqnarray}
V&=& -\mu^2 (H^\dagger H ) + \lambda (H^\dagger H)^2 - \mu_A^2 (S^\dagger S ) + \lambda_1 (S^\dagger S)^2\nonumber \\
&&  +  \lambda_2 (H^\dagger H) (S^\dagger S ) - {1\over 2} \mu_B^{2} S^2 + {1\over 2 } \lambda_3  S^4 + {\rm h.c.} \label{potential}
\end{eqnarray}
where all couplings are real. 
Due to the existence of the last two terms in (\ref{potential}), the CP can be broken spontaneously. 

By setting $H=(0,~v+h)^T/\sqrt{2}$ and $S=(v_s e^{i \varphi} + \alpha + i \beta )/\sqrt{2}$, where $v$ and $v_s$ are VEVs  and  $\varphi$  is the CP phase, the tadpole conditions have three solutions at the zero temperature as follows: (I) $v_s\neq 0$, $v\neq 0$ and $\varphi\neq 0$, (II) $v\neq 0$ $v_s=\varphi=0$ and (III) $v=0$, $v_s\neq0$, $\varphi \neq 0$, depending on parameter settings of the potential.
It is obvious that the third solution is not real at the zero temperature. 
If  the solution (I) is true, there  will be the mixing between the SM Higgs and the scalar singlet, which is strongly constrained by the  Higgs data from the LHC and low energy precision observables.  
This scenario turns to be less attractive.  
We are interested in the solution (II) at the zero temperature, where there is no mixing between the SM Higgs and extra scalars.   
Conditions for this solution being the global minimum is
\begin{eqnarray}
 \lambda_1 > \lambda_3  \hspace{2cm}\mu_A^2 < {1\over 2 } \lambda_2 v^2  \; . \label{constraint}
\end{eqnarray}
For this case scalar masses can be written as  $m_h^2 = 2 \lambda v^2 $, $m_\alpha^2 = -\mu_A^2 -\mu_B^2 + {1 \over 2 } \lambda_2 v^2 $ and $m_\beta^2 =-\mu_A^2 + \mu_B^2 + {1\over 2 } \lambda_2 v^2 $.  
If there is no extra  interactions for the scalar singlet, the lighter component of $S$ will be stable dark matter candidate~\cite{Cline:2012hg}. 
The situation can be different at the finite temperature, because the following thermal mass corrections should be included in the potential:
\begin{eqnarray}
&&\Pi_h\equiv D_h T^2 =\left\{ { 3 g^2 + g^{\prime 2 } \over 16} + {\lambda \over 2 } + {h_t^2 \over 4 } + {\lambda_2 \over 12 } \right\} T^2  \; , \\
&&\Pi_\alpha =\Pi_\beta \equiv D_s T^2= \left( {\lambda_1 \over 3} + {\lambda_2 \over 6 } \right) T^2 \; ,
\end{eqnarray}
where $h_t $ is the top quark Yukawa coupling, $g$ and $g^\prime$ are the $SU(2)_L$ and $U(1)_Y$ couplings.  
$\Pi_h$, $\Pi_\alpha $ and $\Pi_\beta $ are thermal masses of $h$, $\alpha$ and $\beta$ respectively.  
Here we neglect corrections from Coleman-Weinberg  terms, diasy contributions and ${\cal O} (T^4)$ corrections for simplicity while without changing inherent physics. 
For this scenario, once upon a time at $T_s$  in the thermal history of the universe,  the Higgs potential first evolves to the minimum at $v=0$ and $|v_s|\neq0$, resulting in a spontaneous CP violation.
The CP phase is
\begin{eqnarray}
\varphi =\pm {1\over 2 } \arccos\left[ {\lambda_1 -\lambda_3 \over 2 \lambda_3 } {m_\beta^2 -m_\alpha^2 \over \lambda_2 v^2 -m_\alpha^2 -m_\beta^2 + 2 \Pi_\alpha}\right] \; .\label{phase}
\end{eqnarray}
As temperature drops lower to the $T_n$, bubbles relating to electroweak symmetry broken phase, i.e. $v\neq0$ and $v_s =0$, starts to nucleate in the  $v=0$ and $v_s\neq0$ background, and the spontaneous CP phase  evolves to null from the outside of the bubble wall to the inside of the bubble wall.  
In short, the spontaneous CP violation emerges at $T \in (T_n,~T_s)$.

 \noindent {\bf  EWPT} 
 To study the strength of the EWPT, one needs the effective potential at the finite temperature in terms of background fields $h$, $\alpha$ and $\beta$
\begin{eqnarray}
\bar V&=& -{1\over 2 } (\mu^2- \Pi_h ) h^2 + {1 \over 4 }\lambda h^4 + {1\over 4 } \lambda_2 h^2 (\alpha^2 + \beta^2 ) \nonumber\\
&&-{1\over 2 } (\mu_A^2 + \mu_B^2 -\Pi_\alpha )\alpha^2 -{1 \over 2 } ( \mu_A^2 -\mu_B^2 -\Pi_{\beta} ) \beta^2\nonumber\\
&& + {1 \over 4 } (\lambda_1+ \lambda_3) (\alpha^4 + \beta^4 )  + {1\over 2 } (\lambda_1 -3 \lambda_3 ) \alpha^2 \beta^2  \; .
\end{eqnarray}
where $\mu^2, ~\mu_A^2,~ \mu_B^2$  and $\lambda$ can be replaced with physical parameters.

For the parameter settings that satisfy eq. (\ref{constraint}), the scalar singlet gets no VEV at the zero temperature, but it may get a VEV at the finite temperature and the CP is spontaneously broken in the meanwhile. 
As a result, there are two minimums at the critical temperature $T_C$, one at $h=0,~ \alpha= v_s c_\varphi$ and $ \beta= v_s s_\varphi $ another one at $h=v_T, ~\alpha=\beta=0$, where $c_\varphi =\cos \varphi$ and $s_\varphi =\sin \varphi$, and there is barrier between these two vacuums at the tree level~\cite{Espinosa:2011ax,Espinosa:2011eu}.
The condition for the electroweak vacuum being the global one at the temperature below $T_C$ is 
\begin{eqnarray}
D_h \sqrt{\lambda_{\varrho} }> D_s \sqrt{\lambda} \label{constraint2}
\end{eqnarray}
where $\lambda_{\varrho}= \lambda_1 +\lambda_3 \cos(4\varphi) $.

The critical temperature can be calculated using the degenerate condition $ V(0, v_s c_\alpha^{}, v_s s^{}_\alpha, T_C) = V(v_T^{} ,0, 0, T_C) \; ,$
and it results in 
\begin{eqnarray}
8 (\lambda_1 \lambda_3 -\lambda_3^2 ) (\lambda v_0^2 - \Pi_h)^2 =  \lambda (\lambda_1 -\lambda_3 ) (m_\beta^2 -m_\alpha^2)^2 &&  \nonumber \\ + 2\lambda \lambda_3 (\lambda_2 v_0^2 -m_\alpha^2 -m_\beta^2-2\Pi_\alpha )^2 && \; . \nonumber
\end{eqnarray}
where $v_0$ is the VEV of the SM Higgs at the zero temperature.
The gauge invariant and temperature dependent VEV of the SM Higgs is
$
\bar v(T)= \sqrt{  v_0^2 - \Pi_h / \lambda }
$, with which one can estimate the strength of the EWPT.
The condition of the strongly first order EWPT is ${ \bar v(T_C )/ T_C} \ge 1 $~\cite{Quiros:1999jp}.

The bubble wall width can be determined by solving the $O(3)$-Euclidean equation of motion for $\phi_i(r)$:
\begin{eqnarray}
{d^2 \phi_i \over d r^2 } + {2\over r } {d \phi_i \over d r } = \bar V^\prime (\vec{\phi}) \; .
\end{eqnarray}
Boundary conditions for the ${\cal O}(3)$ solution are: $\phi_1 =v_c$, $v_2=v_3 =0$,~$\phi_i^\prime=0$ at $r=0$ and  $\phi_1=0$, $\phi_2=v_s c_\alpha$,~$v_3=v_s s_\alpha$ at $r=\infty$.
Following the technique developed in Ref.~\cite{Espinosa:2011ax,Espinosa:2011eu}, one can analytically determine the bubble wall width in the thin-wall approximation, 
\begin{eqnarray}
L_w^2 &\approx& 1.35 { \lambda + \sqrt{\lambda \lambda_{\varrho}} \over (\lambda_2 -2 \sqrt{\lambda \lambda_{\varrho}}) [\lambda v_0^2 -\Pi_h (T_C^2)]} \nonumber \\
&\times& \left(  1+ \sqrt{ \lambda^2_2 \over 4 \lambda \lambda_{\varrho}}~\right) \; . \label{wallwidth}
\end{eqnarray}
It can be used to estimate the free energy arisen from the bubble wall, which will be done in the next section.
It should be mentioned that eq. (\ref{wallwidth}) is invalided for some extreme scenarios, for example  when $\lambda_2 -2 \sqrt{\lambda \lambda_{\varrho}}\to 0$,  where the bubble wall width can only be calculated numerically with the shooting method.

\begin{figure}[t!]
\centering
\includegraphics[width=0.22\textwidth]{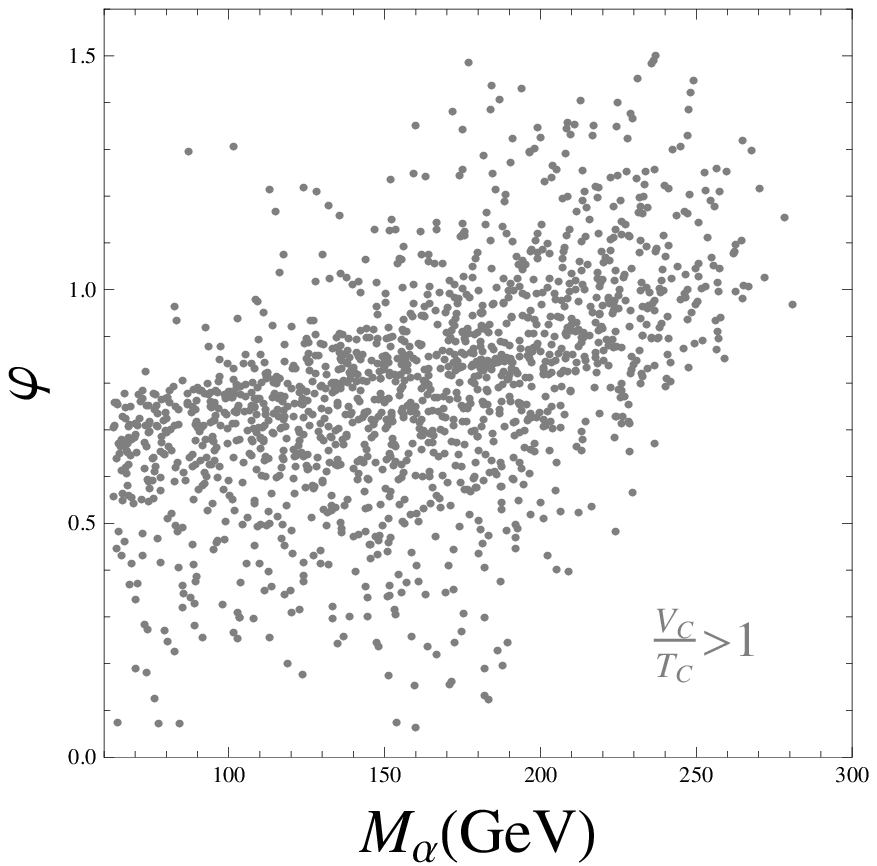}
\includegraphics[width=0.225\textwidth]{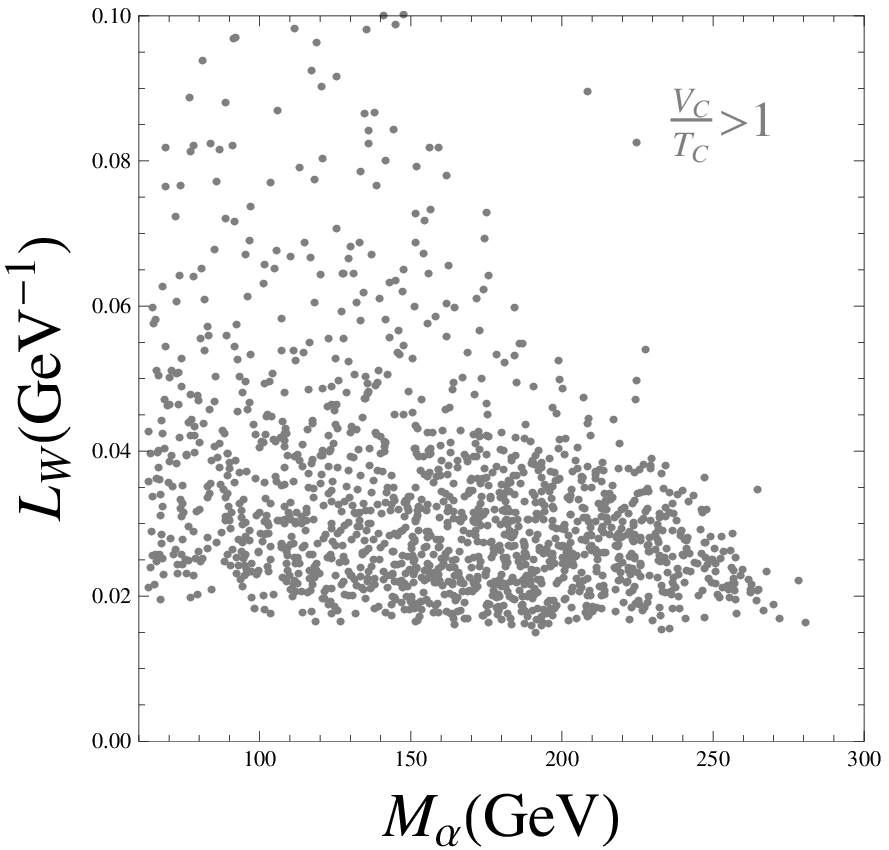}
\caption{
Scattering plots of the CP phase (left-panel) and bubble wall width (right-panel) as the function of the mass of the CP-even scalar singlet. For parameter settings see the text. 
 }\label{phasetst}
\end{figure}

We plot in the left-panel of the Fig.~\ref{phasetst} the spontaneous CP phase as the function of $M_\alpha$, which is the mass of the CP-even scalar singlet,   by setting $\lambda_{1,2,3}$ as random values varying in $(0, ~2)$, which satisfy eqs. (\ref{constraint}) and (\ref{constraint2}) as well as $v_C/T_C \geq 1$.    
One can see that the CP phase approaches to $\pi/2$ with the increase of the $M_\alpha$. 
We show in the right panel of the  Fig.~\ref{phasetst} the bubble wall width as the function of $M_\alpha$,  using the same parameter settings, which shows that $L_w\sim {\cal O} (1)/T_C^{}$. 
For the gravitational waves generated from the EWPT and their signals in the space based interferometer, i.e. LISA , BBO and Tianqin, we refer the reader to Ref.~\cite{Chao:2017vrq}  for detail.  

%%%%%%%%%%%%%%%%%%%%%%%%%%%%%%%%%%%%%%%%%%%%%%%%%%%%
%%%%%%%%%%%%%%%%%%%%%%%%%%%%%%%%%%%%%%%%%%%%%%%%%%%%

 \noindent {\bf  EWBG} 
To generate nonzero baryon asymmetry during the electroweak phase transition one needs a  CP-violating source that leads to the production of  non-zero number densities of fermions near the expanding bubble wall. 
Due to the existence of the Yukawa interaction in eq. (\ref{modell}), the scattering of the right-handed top quark off the bubble wall produce a non-zero number density, which is diffused into the plasma and is translated into the non-zero number density of $n_L$ via the inelastic scattering.  
The weak sphaleron process biases the baryon density against the anti-baryon density, resulting in a non-zero baryon asymmetry which is eaten by the expanding bubble, inside which sphalerons are inactive,  and  is then restored.

The CP-violating source term can be calculated using the ``VEV insertion" method~\cite{Lee:2004we,Riotto:1998zb,Chung:2009qs},
\begin{eqnarray}
S_{\rm top}^{\rm CPV} = -2 \zeta^2  v_s^2 \dot{\varphi} \int { k^2 dk \over \pi^2 \omega_L \omega_R } {\rm Im}\left \{  (\varepsilon_L \varepsilon_R^* -k^2 )  \right.  &&\nonumber \\  { n(\varepsilon_L) - n(\varepsilon_R^* ) \over (\varepsilon_L - \varepsilon_R^* )^2} + \left. (\varepsilon_L \varepsilon_R + k^2 ) { n(\varepsilon_L ) + n(\varepsilon_R) \over (\varepsilon_L + \varepsilon_R)^2 }\right\}&&
\end{eqnarray}  
where the dot represents derivative with respect to $\bar z$ with $\bar z= z+ v_w t$; $n(x)$ is the Fermi distribution; $\varepsilon_{L,R}=\omega_{\mathfrak{t}_L,t_R} -i \Gamma_{\mathfrak{t}_L,t_R}$ are complex poles of the spectral function.
The Transport equations  can be written as
\begin{eqnarray}
\partial^\mu Q_\mu &=& +\Gamma_{m_t} {\cal R}_T^- + \Gamma_{Y_t} \delta_t +\Gamma_{y^\prime} \delta_{\mathfrak{t}^\prime} + 2 \Gamma_{s} \delta_s \nonumber \\
\partial^\mu T_\mu &=&- \Gamma_{m_t}{\cal R}_T^- - \Gamma_{Y_t} \delta_t  -\Gamma_{s} \delta_s -\Gamma_{\zeta} \delta_\mathfrak{t}  \nonumber \\ 
&&+ \Gamma^+_{\mathfrak{t}} {\cal R}_\mathfrak{t}^+ +\Gamma^-_{\mathfrak{t}} {\cal R}_\mathfrak{t}^- +  S_{\rm top}^{\rm CPV} \nonumber \\
\partial^\mu \mathfrak{t}_\mu &=&+\Gamma_{m_\mathfrak{t}}  {\cal R}_\Lambda^-- \Gamma^+_{\mathfrak{t}} {\cal R}_\mathfrak{t}^+ -\Gamma^-_{\mathfrak{t}}{\cal R}_\mathfrak{t}^- + \Gamma_\zeta \delta_\mathfrak{t} - S_{\rm top}^{\rm CPV} \nonumber \\
\partial^\mu \mathfrak{t}^\prime_\mu &=&-\Gamma_{m_\mathfrak{t}}  {\cal R}_\Lambda^--\Gamma_{y^\prime} \delta_{\mathfrak{t}^\prime}  \nonumber \\
\partial^\mu S_\mu &=& -\Gamma_\zeta \delta_\mathfrak{t} \nonumber \\
\partial^\mu H_\mu &=& - \Gamma_{Y_t} \delta_t -\Gamma_{y^\prime} \delta_{\mathfrak{t}^\prime} 
\end{eqnarray}
where $Q $, $T$, $\mathfrak{t}$ and $\mathfrak{t}'$ are number densities of $Q_{3 L}$, $t_R$, $\mathfrak{t}_L$ and $\mathfrak{t}_R$ respectively; $\partial^\mu =v_w {d /d \bar z} -D_a {d^2 / d \bar z^2}$ with $D_a$  the diffusion constant and $v_w$ the bubble wall velocity; ${\cal R}_T^- \equiv \left( { T /  k_T} - {Q / k_Q}\right)$, ${\cal R}_\mathfrak{t}^\pm = T/k_T \pm \mathfrak{t}/k_\mathfrak{t} $, $\delta_s\equiv\left(  {T / k_T } -2 {Q / k_Q } + 9 { B /k_B }\right) $, $ \delta_t\equiv \left( {T / k_T} - {Q / k_Q} +{H/ k_H }  \right) $, $ \delta_{\mathfrak{t}^\prime} \equiv \left( { \mathfrak{t}^\prime/ k_\mathfrak{t}} - {Q / k_Q} +{H/ k_H }  \right) $, ${\cal R}_\Lambda^-=(\mathfrak{t}^\prime/k_\mathfrak{t}-\mathfrak{t}/k_\mathfrak{t})$; $n_i$ and $k_i$ are the number density and the statistical factor for the particle $``i"$, respectively.
Coefficients $\Gamma_{Y_t}$, $\Gamma_{y^\prime}$ and $\Gamma_\zeta$   denotes the interaction rates arising from the SM top quark Yukawa interaction and the Yukawa interaction in Eq. (2), respectively; $\Gamma_{m_{\mathfrak{t}}}$, $\Gamma^{}_{m_t}$ and  $\Gamma_\mathfrak{t}^\pm$ are the CP-conserving scattering rates of  particles with the background Higgs field on the bubble wall; $\Gamma_{s} \equiv 16\kappa^\prime  \alpha_s^4 T$ is the strong sphaleron rate with $\alpha_s$ the strong coupling and $\kappa^\prime\sim{\cal O}(1)$. 

Transport equations can be solved numerically using the relaxation method. 
The final BAU can be written in the term of integral of the left-handed fermion charge density, which is $n_L(z) = 5 Q + 4 T $,
\begin{eqnarray}
\hat n_B = - { 3 \Gamma_{ws} \over 2 D_Q \lambda_+} \int_{-\infty}^{-L_w/2} d z n_L (z) e^{- \lambda_- z } \nonumber 
\end{eqnarray} 
where $\lambda_\pm ={1 \over 2 D_Q } (v_w \pm \sqrt{v_w^2 + 4 D_Q R })$,  $R\sim 2\times 10^{-3} ~{\rm GeV}$~\cite{Chung:2009cb} being the inverse washout factor for the electroweak sphaleron transitions.  
The weak sphaleron rate is $\Gamma_{ws} =6\kappa \alpha_w^5 T$, where $\kappa \approx 20$~\cite{DOnofrio:2014rug} and $\alpha_w =g_2^2/4\pi \approx 1/30$.

Notice that  the solution for the spontaneous CP  violating phase given in eq. (\ref{phase}) have a reflect symmetry : $\varphi\to - \varphi$, under which the free energy of bubble is invariant.
As a result, there are two different kinds of bubbles relating to two possible signs of $\varphi$,  which create baryons of opposite signs. 
The net baryon asymmetry averaged over the entire universe will be zero resulting in a null BAU, since there is no preferred bubble against the other one.  
There are two alternative ways to avoid this problem:
\begin{itemize}
\item Adding soft $Z_2$-symmetry broken terms,  i.e. $\kappa S+{\rm h.c.}$ and/or $\eta S^3+ {\rm h.c.}$, to the potential;
\item Introducing a small explicit CP violating phase in the Higgs potential: $\mu_B^2 \to \mu_B^2 e^{ i \delta}$.
\end{itemize}
 As a result, the degeneracy between two types of  vacuums, $+\varphi$ and $-\varphi$,  will be broken during the phase transition.  And the false vacuum with higher free energy will decay into the true vacuum with lower free-energy via the bubble nucleation at a temperature $T_G^{}$.   
 If $T_G>T_n$, where $T_n$ is the electroweak bubble nucleation temperature, there will be only one kind of  phase left when the second step phase transition takes place, and  one has net BAU produced via the conventional EWBG. 
 One the other hand,  if $T_G< T_n$,  vacuums with $\pm \varphi$ are still there equally at the time of the second step phase transition, but volumes occupied by electroweak bubbles raised on the $\pm \varphi$ background  can be significantly different. 
 The ratio between two kinds of bubbles takes the form~\cite{Comelli:1993ne}
\begin{eqnarray}
{N_+\over N_-} = \exp \left(\Delta F \over T \right)
\end{eqnarray}
where $N_{\pm}$ are the number densities of bubble with $\pm \varphi$, $\Delta F $ is the difference of the free energy between two types of bubbles.  
The free energy of a  vacuum have two terms:  the surface term $F_S$ and the volume term $F_V$. 
For our case $\Delta F $ mainly come from the difference of surface terms, which can be evaluated analytically~\cite{Anderson:1991zb}.
%\begin{eqnarray}
%F_S = 4\pi \int_{R-L_w}^{R+ L_w } r^2 d r \left ( {1 \over 2 } \sum_i \left(d \phi_i\over d r \right)^2  + V( \vec \phi) \right)
%\end{eqnarray}
Finally, the global baryon density is written in term of asymmetry of two kinds of bubbles,
\begin{eqnarray}
n_B =\hat n_B^{(+)}  {N_+-N_- \over N_+ + N_- }  \; ,
\end{eqnarray}
where $\hat n_B^{(+)}$ is the BAU generated from the $+\varphi$ bubble. 
In the limit $\Delta F\to \pm \infty$, it has $n_B \to \hat n_B^{\pm}$.
For simplicity, we add $\Delta S + {\rm h.c.}$ term to the effective potential, where $\Delta$ is a small real coupling,  and the Higgs potential contains no explicit CP violation.

%%%%%%%%%%%%%%%%%%%%%%%%%%%%%%%%%%%%%%%%%%%%%%%%%%%%
\begin{figure}[t!]
\centering
\includegraphics[width=0.23\textwidth]{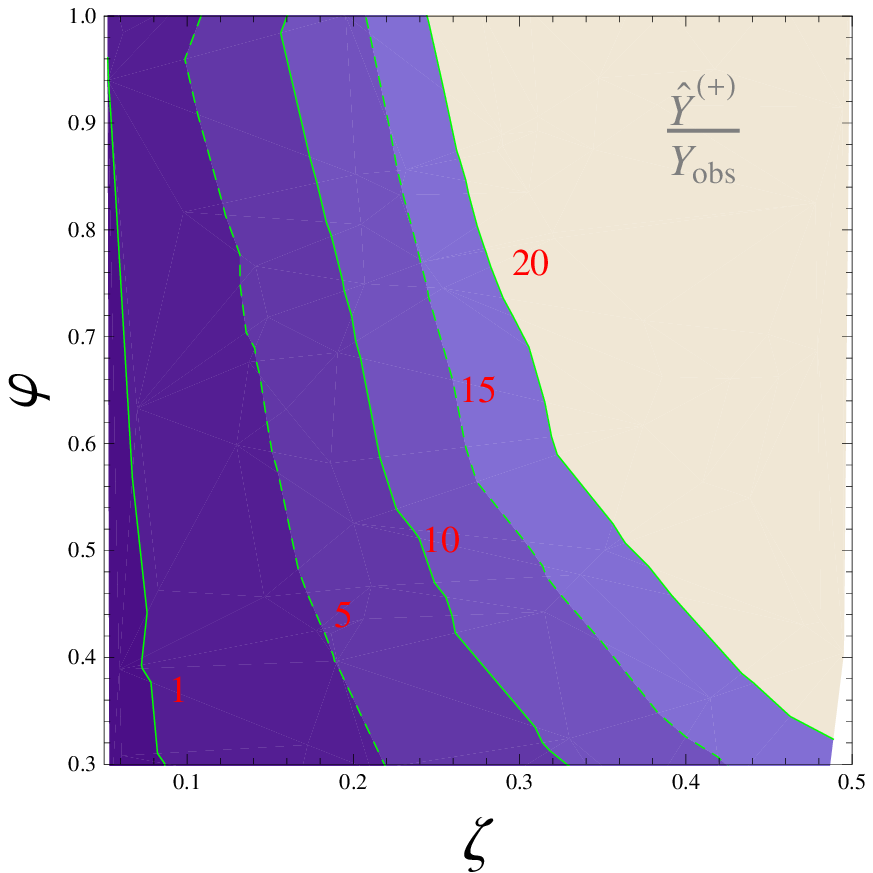}
\includegraphics[width=0.23\textwidth]{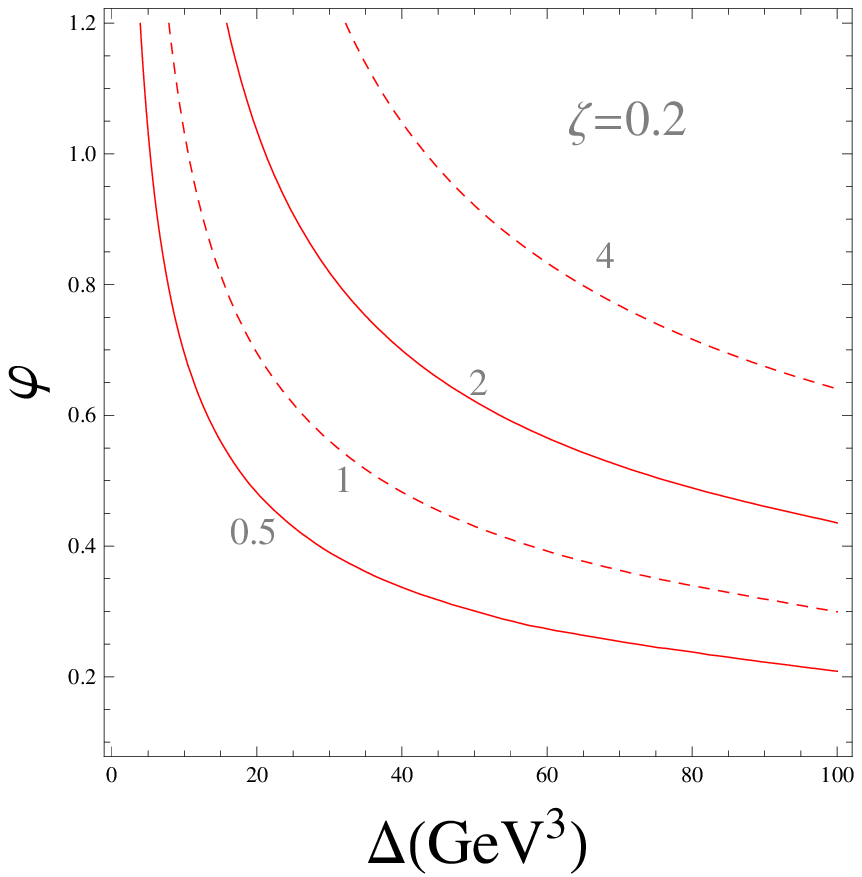}
\caption{
Left panel: Contours of $\hat Y^{(+)}$ in units of $Y_{\rm obs}$  in the $\zeta-\varphi$ plane; Right panel: Contours of $Y_B/Y_{\rm obs}$ in the $\Delta-\varphi$ plane by setting $\zeta=0.2$.
 }\label{fig-bau}
\end{figure}
%%%%%%%%%%%%%%%%%%%%%%%%%%%%%%%%%%%%%%%%%%%%%%%%%%%%

%
As an illustration, we show in the left-panel of the Fig.~\ref{fig-bau} contours of $\hat Y^{(+)}$, which is the baryon number density generated by the bubble with $+\varphi$, divided by  the observed value in the  $\zeta-\varphi$ plane, by setting $T_C\sim100~{\rm GeV}$, $M_\mathfrak{t}=500~{\rm GeV}$, $y^\prime=0$, $\Lambda=1~{\rm TeV}$ and $\Gamma_{\mathfrak{t}_R}=5\zeta^2$. 
The Higgs VEV profiles are taken from Refs.~\cite{Cline:2012hg,Cline:2017qpe,Espinosa:2011eu}, using  the bubble wall width given in eq. (\ref{wallwidth}).  
Since the bubble wall velocity is relevant to the friction in the wall, which is still unclear,  we set $v_w=0.2$ for simplicity. 
Transport equations are solved numerically with the ``relaxation" method~\cite{recipes}.
Compared with the EWBG in the MSSM, the CP source term  in the MSSM is proportional to a small value $\Delta \beta$, which is usually taken as $0.01$~\cite{Chung:2009qs},  while the CP source in our case is proportional to the spontaneous CP phase $\varphi$, which is taken as $0.8$.
It explains why the observed BAU can be generated with a tiny $\zeta$ in our case.  
We show in the right-panel of the Fig.~\ref{fig-bau} contours of the BAU in units of $Y_{\rm obs}$ in the $\Delta-\varphi$ plane by fixing $\zeta=0.2$. Other inputs are the same as above.
Apparently the BAU can be explained by a small $|\Delta|$ in our scenario.
It should be mentioned that further calculations are needed to determine the temperature of domain wall collapsing  as well as the friction parameter relating to the bubble expansion. 
We defer them to a another work.

 \noindent {\bf  Summary and discussion} 
 Extending the SM with a complex scalar singlet and a vector-like top quark, we showed that  our universe may undergo a two-step phase transition, and the  spontaneous CP phase can emerge between the first-step and the second-step phase transitions. 
The CP violation can be responsible for the origin of the BAU  if a small $Z_2$ breaking term  is added to the Higgs potential, which causes the collapse of the domain wall, while no extra  CP violation is needed. 
Since the spontaneous CP phase  disappears after the electroweak phase transition, there is no EDM constraint on the model,  and the tension between the non-observation of EDMs and the requirement of a large CP violation by the EWBG is released. 
It opens a new window for studies of CP violation as well as EWBG.

The BAU is generated from interactions of vector-like top quark in our model.  Actually it may also arise from a dimension-5 effective operator~\cite{Espinosa:2011eu}, or a dimension-6 effective operator~\cite{Cline:2012hg}.
Integrating out the vector-like quark in eq. (2), one gets the dimension-5 effective operator given in Ref.~\cite{Espinosa:2011eu}.
For our case, $S$ can be two-components top-flavored as well as Higgs portal  dark matter candidates in the limit $y^\prime \to 0$. 
Collider signatures of $S$ will be Di-Higgs in various channels or di-jet plus missing energy. 
We refer the reader to Ref.~\cite{Curtin:2014jma,Chen:2017qcz} for collider searches of  the  first order EWPT, which is in common with our case.
To distinguish our model, one needs to determine at least  two types of Higgs interactions at colliders, i.e. $h\alpha \alpha$ and $h\beta \beta$, whose signatures at the LHC and CEPC will be studied in a longer paper.

 \noindent {\bf  Acknowledgements} I  am indebted to Michael J. Ramsey-Musolf  for his support at the UMASS-Amherst and  frequent discussions; to  Huai-ke Guo for his help on numerical simulations. We thank Jonathan Kozaczuk and Stefano Profumo for illuminating discussions.


\begin{thebibliography}{99}


%\cite{Ade:2015xua}
\bibitem{Ade:2015xua} 
  P.~A.~R.~Ade {\it et al.} [Planck Collaboration],
  %``Planck 2015 results. XIII. Cosmological parameters,''
  Astron.\ Astrophys.\  {\bf 594}, A13 (2016)
  doi:10.1051/0004-6361/201525830
  [arXiv:1502.01589 [astro-ph.CO]].
  %%CITATION = doi:10.1051/0004-6361/201525830;%%
  %3286 citations counted in INSPIRE as of 09 May 2017


%\cite{Sakharov:1967dj}
\bibitem{Sakharov:1967dj} 
  A.~D.~Sakharov,
  %``Violation of CP Invariance, c Asymmetry, and Baryon Asymmetry of the Universe,''
  Pisma Zh.\ Eksp.\ Teor.\ Fiz.\  {\bf 5}, 32 (1967)
  [JETP Lett.\  {\bf 5}, 24 (1967)]
  [Sov.\ Phys.\ Usp.\  {\bf 34}, 392 (1991)]
  [Usp.\ Fiz.\ Nauk {\bf 161}, 61 (1991)].
  doi:10.1070/PU1991v034n05ABEH002497
  %%CITATION = doi:10.1070/PU1991v034n05ABEH002497;%%
  %2653 citations counted in INSPIRE as of 10 May 2017



%\cite{Huet:1994jb}
\bibitem{Huet:1994jb} 
  P.~Huet and E.~Sather,
  %``Electroweak baryogenesis and standard model CP violation,''
  Phys.\ Rev.\ D {\bf 51}, 379 (1995)
  doi:10.1103/PhysRevD.51.379
  [hep-ph/9404302].
  %%CITATION = doi:10.1103/PhysRevD.51.379;%%
  %318 citations counted in INSPIRE as of 11 May 2017
  
  %\cite{Kuzmin:1985mm,Rubakov:1996vz}
\bibitem{Kuzmin:1985mm} 
  V.~A.~Kuzmin, V.~A.~Rubakov and M.~E.~Shaposhnikov,
  %``On the Anomalous Electroweak Baryon Number Nonconservation in the Early Universe,''
  Phys.\ Lett.\  {\bf 155B}, 36 (1985).
  doi:10.1016/0370-2693(85)91028-7
  %%CITATION = doi:10.1016/0370-2693(85)91028-7;%%
  %2367 citations counted in INSPIRE as of 17 May 2017
  
  %\cite{Rubakov:1996vz}
\bibitem{Rubakov:1996vz} 
  V.~A.~Rubakov and M.~E.~Shaposhnikov,
  %``Electroweak baryon number nonconservation in the early universe and in high-energy collisions,''
  Usp.\ Fiz.\ Nauk {\bf 166}, 493 (1996)
  [Phys.\ Usp.\  {\bf 39}, 461 (1996)]
  doi:10.1070/PU1996v039n05ABEH000145
  [hep-ph/9603208].
  %%CITATION = doi:10.1070/PU1996v039n05ABEH000145;%%
  %704 citations counted in INSPIRE as of 17 May 2017

%\cite{Morrissey:2012db}
\bibitem{Morrissey:2012db} 
  D.~E.~Morrissey and M.~J.~Ramsey-Musolf,
  %``Electroweak baryogenesis,''
  New J.\ Phys.\  {\bf 14}, 125003 (2012)
  doi:10.1088/1367-2630/14/12/125003
  [arXiv:1206.2942 [hep-ph]].
  %%CITATION = doi:10.1088/1367-2630/14/12/125003;%%
  %212 citations counted in INSPIRE as of 17 May 2017
  
 
 %\cite{Gabbiani:1996hi}
\bibitem{Gabbiani:1996hi} 
  F.~Gabbiani, E.~Gabrielli, A.~Masiero and L.~Silvestrini,
  %``A Complete analysis of FCNC and CP constraints in general SUSY extensions of the standard model,''
  Nucl.\ Phys.\ B {\bf 477}, 321 (1996)
  doi:10.1016/0550-3213(96)00390-2
  [hep-ph/9604387].
  %%CITATION = doi:10.1016/0550-3213(96)00390-2;%%
  %1244 citations counted in INSPIRE as of 24 May 2017

%\cite{Inoue:2014nva}
\bibitem{Inoue:2014nva} 
  S.~Inoue, M.~J.~Ramsey-Musolf and Y.~Zhang,
  %``CP-violating phenomenology of flavor conserving two Higgs doublet models,''
  Phys.\ Rev.\ D {\bf 89}, no. 11, 115023 (2014)
  doi:10.1103/PhysRevD.89.115023
  [arXiv:1403.4257 [hep-ph]].
  %%CITATION = doi:10.1103/PhysRevD.89.115023;%%
  %57 citations counted in INSPIRE as of 24 May 2017
  
 
  
  %\cite{Lee:2004we}
\bibitem{Lee:2004we} 
  C.~Lee, V.~Cirigliano and M.~J.~Ramsey-Musolf,
  %``Resonant relaxation in electroweak baryogenesis,''
  Phys.\ Rev.\ D {\bf 71}, 075010 (2005)
  doi:10.1103/PhysRevD.71.075010
  [hep-ph/0412354].
  %%CITATION = doi:10.1103/PhysRevD.71.075010;%%
  %88 citations counted in INSPIRE as of 18 May 2017
  
  %\cite{Chao:2016aer}
\bibitem{Chao:2016aer} 
  W.~Chao,
  %``The Diphoton Excess Inspired Electroweak Baryogenesis,''
  arXiv:1601.04678 [hep-ph].
  %%CITATION = ARXIV:1601.04678;%%
  %40 citations counted in INSPIRE as of 18 May 2017
  
  %\cite{Chao:2014dpa}
\bibitem{Chao:2014dpa} 
  W.~Chao and M.~J.~Ramsey-Musolf,
  %``Electroweak Baryogenesis, Electric Dipole Moments, and Higgs Diphoton Decays,''
  JHEP {\bf 1410}, 180 (2014)
  doi:10.1007/JHEP10(2014)180
  [arXiv:1406.0517 [hep-ph]].
  %%CITATION = doi:10.1007/JHEP10(2014)180;%%
  %11 citations counted in INSPIRE as of 18 May 2017
  
  %\cite{Chao:2015uoa}
\bibitem{Chao:2015uoa} 
  W.~Chao and M.~J.~Ramsey-Musolf,
  %``Catalysis of Electroweak Baryogenesis via Fermionic Higgs Portal Dark Matter,''
  arXiv:1503.00028 [hep-ph].
  %%CITATION = ARXIV:1503.00028;%%
  %4 citations counted in INSPIRE as of 18 May 2017

%\cite{Guo:2016ixx}
\bibitem{Guo:2016ixx} 
  H.~K.~Guo, Y.~Y.~Li, T.~Liu, M.~Ramsey-Musolf and J.~Shu,
  %``Lepton-Flavored Electroweak Baryogenesis,''
  arXiv:1609.09849 [hep-ph].
  %%CITATION = ARXIV:1609.09849;%%
  %3 citations counted in INSPIRE as of 18 May 2017



%\cite{Li:2008kz}
\bibitem{Li:2008kz} 
  Y.~Li, S.~Profumo and M.~Ramsey-Musolf,
  %``Higgs-Higgsino-Gaugino Induced Two Loop Electric Dipole Moments,''
  Phys.\ Rev.\ D {\bf 78}, 075009 (2008)
  doi:10.1103/PhysRevD.78.075009
  [arXiv:0806.2693 [hep-ph]].
  %%CITATION = doi:10.1103/PhysRevD.78.075009;%%
  %52 citations counted in INSPIRE as of 24 May 2017

 %\cite{Baron:2013eja}
\bibitem{Baron:2013eja} 
  J.~Baron {\it et al.} [ACME Collaboration],
  %``Order of Magnitude Smaller Limit on the Electric Dipole Moment of the Electron,''
  Science {\bf 343}, 269 (2014)
  doi:10.1126/science.1248213
  [arXiv:1310.7534 [physics.atom-ph]].
  %%CITATION = doi:10.1126/science.1248213;%%
  %335 citations counted in INSPIRE as of 18 May 2017
 
 
  %\cite{Espinosa:2011ax}
\bibitem{Espinosa:2011ax} 
  J.~R.~Espinosa, T.~Konstandin and F.~Riva,
  %``Strong Electroweak Phase Transitions in the Standard Model with a Singlet,''
  Nucl.\ Phys.\ B {\bf 854}, 592 (2012)
  doi:10.1016/j.nuclphysb.2011.09.010
  [arXiv:1107.5441 [hep-ph]].
  %%CITATION = doi:10.1016/j.nuclphysb.2011.09.010;%%
  %107 citations counted in INSPIRE as of 28 Apr 2017
  
  %\cite{Espinosa:2011eu}
\bibitem{Espinosa:2011eu} 
  J.~R.~Espinosa, B.~Gripaios, T.~Konstandin and F.~Riva,
  %``Electroweak Baryogenesis in Non-minimal Composite Higgs Models,''
  JCAP {\bf 1201}, 012 (2012)
  doi:10.1088/1475-7516/2012/01/012
  [arXiv:1110.2876 [hep-ph]].
  %%CITATION = doi:10.1088/1475-7516/2012/01/012;%%
  %40 citations counted in INSPIRE as of 28 Apr 2017
 

  
 
 
 %\cite{Haber:2012np}
\bibitem{Haber:2012np} 
  H.~E.~Haber and Z.~Surujon,
  %``A Group-theoretic Condition for Spontaneous CP Violation,''
  Phys.\ Rev.\ D {\bf 86}, 075007 (2012)
  [arXiv:1201.1730 [hep-ph]].
  %%CITATION = ARXIV:1201.1730;%%
  %3 citations counted in INSPIRE as of 22 Nov 2014
  
  
  %\cite{Quiros:1999jp}
\bibitem{Quiros:1999jp} 
  M.~Quiros,
  %``Finite temperature field theory and phase transitions,''
  hep-ph/9901312.
  %%CITATION = HEP-PH/9901312;%%
  %190 citations counted in INSPIRE as of 24 May 2017

   %\cite{Chao:2017vrq}
\bibitem{Chao:2017vrq} 
  W.~Chao, H.~K.~Guo and J.~Shu,
  %``Gravitational Wave Signals of Electroweak Phase Transition Triggered by Dark Matter,''
  arXiv:1702.02698 [hep-ph].
  %%CITATION = ARXIV:1702.02698;%%
  %7 citations counted in INSPIRE as of 30 Apr 2017

    %\cite{Lee:2004we,Riotto:1998zb,Chung:2009qs}
\bibitem{Riotto:1998zb} 
  A.~Riotto,
  %``The More relaxed supersymmetric electroweak baryogenesis,''
  Phys.\ Rev.\ D {\bf 58}, 095009 (1998)
  doi:10.1103/PhysRevD.58.095009
  [hep-ph/9803357].
  %%CITATION = doi:10.1103/PhysRevD.58.095009;%%
  %100 citations counted in INSPIRE as of 23 May 2017
  
  
  %\cite{Chung:2009qs}
\bibitem{Chung:2009qs} 
  D.~J.~H.~Chung, B.~Garbrecht, M.~J.~Ramsey-Musolf and S.~Tulin,
  %``Supergauge interactions and electroweak baryogenesis,''
  JHEP {\bf 0912}, 067 (2009)
  doi:10.1088/1126-6708/2009/12/067
  [arXiv:0908.2187 [hep-ph]].
  %%CITATION = doi:10.1088/1126-6708/2009/12/067;%%
  %40 citations counted in INSPIRE as of 23 May 2017

  
  %\cite{Chung:2009cb}
\bibitem{Chung:2009cb} 
  D.~J.~H.~Chung, B.~Garbrecht, M.~J.~Ramsey-Musolf and S.~Tulin,
  %``Lepton-mediated electroweak baryogenesis,''
  Phys.\ Rev.\ D {\bf 81}, 063506 (2010)
  doi:10.1103/PhysRevD.81.063506
  [arXiv:0905.4509 [hep-ph]].
  %%CITATION = doi:10.1103/PhysRevD.81.063506;%%
  %24 citations counted in INSPIRE as of 24 May 2017
  
    %\cite{DOnofrio:2014rug}
\bibitem{DOnofrio:2014rug} 
  M.~D'Onofrio, K.~Rummukainen and A.~Tranberg,
  %``Sphaleron Rate in the Minimal Standard Model,''
  Phys.\ Rev.\ Lett.\  {\bf 113}, no. 14, 141602 (2014)
  doi:10.1103/PhysRevLett.113.141602
  [arXiv:1404.3565 [hep-ph]].
  %%CITATION = doi:10.1103/PhysRevLett.113.141602;%%
  %56 citations counted in INSPIRE as of 27 Apr 2017
  


  %\cite{Comelli:1993ne}
\bibitem{Comelli:1993ne} 
  D.~Comelli, M.~Pietroni and A.~Riotto,
  %``Spontaneous CP violation and baryogenesis in the minimal supersymmetric Standard Model,''
  Nucl.\ Phys.\ B {\bf 412}, 441 (1994)
  [hep-ph/9304267].
  %%CITATION = HEP-PH/9304267;%%
  %52 citations counted in INSPIRE as of 01 Dec 2014
  
  
  %\cite{Anderson:1991zb}
\bibitem{Anderson:1991zb} 
  G.~W.~Anderson and L.~J.~Hall,
  %``The Electroweak phase transition and baryogenesis,''
  Phys.\ Rev.\ D {\bf 45}, 2685 (1992).
  %%CITATION = PHRVA,D45,2685;%%
  %286 citations counted in INSPIRE as of 01 Dec 2014
  

  
  %\cite{Cline:2012hg}
\bibitem{Cline:2012hg} 
  J.~M.~Cline and K.~Kainulainen,
  %``Electroweak baryogenesis and dark matter from a singlet Higgs,''
  JCAP {\bf 1301}, 012 (2013)
  [arXiv:1210.4196 [hep-ph]].
  %%CITATION = ARXIV:1210.4196;%%
  %27 citations counted in INSPIRE as of 22 Nov 2014
  
  
  


  
  
  

%\cite{Cline:2012hg,Cline:2017qpe,Espinosa:2011eu}
\bibitem{Cline:2017qpe} 
  J.~M.~Cline, K.~Kainulainen and D.~Tucker-Smith,
  %``Electroweak baryogenesis from a dark sector,''
  arXiv:1702.08909 [hep-ph].
  %%CITATION = ARXIV:1702.08909;%%
  %2 citations counted in INSPIRE as of 23 May 2017
  
  \bibitem{recipes}

W.H. Press, S.A. Teukolsky, W.T. Vetterling and B.P. Flannery, Numerical recipes in C, 2nd edition, Cambridge University Press, Cambridge U.K. (1992).


  
  %\cite{Curtin:2014jma}
\bibitem{Curtin:2014jma} 
  D.~Curtin, P.~Meade and C.~T.~Yu,
  %``Testing Electroweak Baryogenesis with Future Colliders,''
  JHEP {\bf 1411}, 127 (2014)
  doi:10.1007/JHEP11(2014)127
  [arXiv:1409.0005 [hep-ph]].
  %%CITATION = doi:10.1007/JHEP11(2014)127;%%
  %67 citations counted in INSPIRE as of 08 May 2017



%\cite{Chen:2017qcz}
\bibitem{Chen:2017qcz} 
  C.~Y.~Chen, J.~Kozaczuk and I.~M.~Lewis,
  %``Non-resonant Collider Signatures of a Singlet-Driven Electroweak Phase Transition,''
  arXiv:1704.05844 [hep-ph].
  %%CITATION = ARXIV:1704.05844;%%
  %1 citations counted in INSPIRE as of 24 May 2017




\end{thebibliography}
\end{document}